\documentclass[aps,prd,nofootinbib,twocolumn,superscriptaddress,showpacs,floatfix]{revtex4}
\usepackage{mathrsfs}
\usepackage{latexsym}
\usepackage{amsmath}
\usepackage{amssymb}
\usepackage{graphicx}
\usepackage{longtable}

\usepackage{bbm}
\usepackage{epsfig}
\usepackage[usenames]{color}
\usepackage[colorlinks=true,citecolor=blue,linkcolor=blue]{hyperref}
\allowdisplaybreaks

\begin{document}

%\begin{CJK*}{GBK}{song}

%\begin{CJK*}{GB}{}

\title{ Phase transition of strongly interacting matter with a chemical potential dependent Polyakov loop potential}

\author{Guo-yun Shao}
\email[Corresponding author: ]{gyshao@mail.xjtu.edu.cn}
\affiliation{Department of Applied Physics, Xi'an Jiaotong
University, Xi'an, 710049, China}
%\email[Corresponding author: ]{gyshao@mail.xjtu.edu.cn}

\author{Zhan-duo Tang}
%\email[Corresponding author: ]{shaogy@pku.edu.cn}
\affiliation{Department of Applied Physics, Xi'an Jiaotong
University, Xi'an, 710049, China}

\author{Massimo Di Toro}
\affiliation{INFN-Laboratori Nazionali del Sud, Via S. Sofia 62,
Catania,  I-95123, Italy}
\affiliation{Physics and Astronomy Dept., University of Catania, Via S. Sofia 64, Catania,  I-95123, Italy}

\author{Maria Colonna}
\affiliation{INFN-Laboratori Nazionali del Sud, Via S. Sofia 62, 
Catania, I-95123, Italy}
\author{Xue-yan Gao}
%\email[Corresponding author: ]{shaogy@pku.edu.cn}
\affiliation{Department of Applied Physics, Xi'an Jiaotong
University, Xi'an, 710049, China}
\author{Ning Gao}
%\email[Corresponding author: ]{shaogy@pku.edu.cn}
\affiliation{Department of Applied Physics, Xi'an Jiaotong
University, Xi'an, 710049, China}
%\author{Shi-jun Mao}
%\affiliation{Department of Applied Physics, Xi'an Jiaotong
%University, Xi'an, 710049, China}
%\author{Y. L. Zhao}
%\affiliation{School of Energy and Power Engineering, Xi'an Jiaotong University, 710049, Xi'an, China}

%\author{M. Colonna}
%\affiliation{INFN-Laboratori Nazionali del Sud, Via S. Sofia 62, I-95123
%Catania, Italy}

%\author{M. Di Toro}
%\affiliation{INFN-Laboratori Nazionali del Sud, Via S. Sofia 62, I-95123
%Catania, Italy}
%\affiliation{Physics and Astronomy Dept., University of Catania, Via S. Sofia 64, I-95123
%Catania, Italy}

%\author{Y. X. Liu}
%\affiliation{Department of Physics and State Key Laboratory of \\
%Nuclear Physics and Technology,
%Peking University, Beijing 100871, China}

%\affiliation{Center of Theoretical Nuclear Physics,\\ National Laboratory of
%Heavy Ion Accelerator, Lanzhou 730000, China}

%\date{\today}

\begin{abstract}
We construct a hadron-quark two-phase model based on the Walecka-quantum hadrodynamics and the improved Polyakov--Nambu--Jona-Lasinio model with an explicit chemical potential dependence of Polyakov-loop potential ($\mu$PNJL model). With respect to the original PNJL  model, the confined-deconfined phase transition is largely affected at low temperature and large chemical potential. Using the  two-phase model, we investigate
the equilibrium transition between hadronic and quark matter at finite chemical potentials and temperatures. The numerical results show that the transition boundaries from nuclear to quark matter move towards smaller chemical potential (lower density)  when the $\mu$-dependent Polyakov loop potential is taken. In
particular, for charge asymmetric matter, we compute the local
asymmetry of $u, d$ quarks  in
the hadron-quark coexisting phase, and
analyse the isospin-relevant observables possibly measurable in heavy-ion collision (HIC) experiments. In general new HIC data on the location and properties of the mixed phase would bring relevant information on the expected chemical potential dependence of the Polyakov Loop contribution.
\end{abstract}


\maketitle

\section{Introduction}
The exploration of QCD phase diagram of strongly interacting matter
and the transition signatures from nuclear to quark-gluon
matter are subjects of great interest  in recent decades. Intensive
searches on high-energy heavy-ion collision (HIC) have been performed in laboratories such as RHIC and LHC, and a near perfect fluid of quark-gluon plasma (QGP) has been created  \cite{Gupta11}. Further experiments to look for the critical endpoint (CEP) and the boundaries of the phase transition are in plan in the next generation facilities such as the second stage of beam energy scan  (BES II) project on RHIC and programs on NICA/FAIR/J-PARC. In particular,  experiments will be performed in the region of high baryon density where a promising observation of the signatures of the phase transformation is being looked forward to. 

Ultimately these phenomena have to be understood in the frame of quantum chromodynamics (QCD). However, in spite of tremendous theoretical and experimental
efforts, the QCD phase diagram has not been unveiled yet
\cite{Braun09,Fukushima11}.  In particular, at finite chemical potential $\mu_B$, the situation is not clear. 
Lattice QCD simulation is a fundamental tool to investigate the
thermodynamics of QCD matter at vanishing and/or small chemical
potential \cite{Karsch01, Karsch02,Kaczmarek05,Allton02,Cheng06,
Aoki09}, but it suffers the sign problem of the fermion
determinant with three colors at finite baryon chemical potential.
Some approximation methods have been proposed to try to overcome
the problem, however, the region of large chemical potential and low temperature
essentially remains inaccessible
\cite{Fodor07,Elia09,Ejiri08,Clark07}.

In addition to the lattice QCD simulation, kinds of
quantum field
 theory approaches and  phenomenological  models,
such as the Dyson-Schwinger equation approach
\cite{Roberts00,Alkofer01,Fischer06,Maris03,Cloet14,Xu15},
the Nambu--Jona-Lasinio (NJL) model \cite{Hatsuda84,
Klevansky92,Hatsuda94,Alkofer96,Buballa05, Rehberg96,
Huang03,Alford08,Chu15,Cao14,Menezes14,Xujun14}, the PNJL model
\cite{Fukushima04,Ratti06,Costa10,Kashiwa08,Fu08,Ghosh15,Dutra13}, the entanglement extended PNJL (EPNJL) model \cite{Sakai10,Sakai12,Restrepo15, Sasaki12, Sasaki11, Gatto11,Ferreira14},
the Polyakov-loop extended quark-meson  (PQM) model
\cite{Schaefer10, Skokov11,Chatterjee12},
have been developed to give a complete description of QCD matter.

Among these models, the PNJL model which takes into account both the chiral dynamics and
(de)confinement effect at high temperature, gives a good reproduction of lattice data at vanishing chemical potential. 
On the other hand, in the original PNJL model, a ``quarkyonic phase"  in which the quarks are confined but the dynamical chiral symmetry is already restored appear at high density and finite $T$ \cite{McLarren2007, Hidaka2008}. In theory, quark deconfinement should also occur at high density. The absence of quark deconfinement at low $T$ and high density in the original PNJL model originates from that the Polyakov-loop potential is extracted from pure Yang-Mills lattice simulation at vanishing $\mu_B$. In the presence of dynamical quarks, the contribution from matter sector and its quantum back-reaction to the glue sector should be included.
This can realized by introducing a flavor and chemical potential dependent  
Polyakov loop potential in the 
functional renormalization group (FRG) approach \cite{Schaefer2007}. With the incorporation of both the matter and glue dynamics, the flavor and chemical potential dependent Polyakov loop potential has been taken in the PQM model \cite{Schaefer2007,Herbst2011,Herbst2013}  and PNJL model \cite{Abuki08,Xin14} to study the full QCD phase diagram and thermodynamics. The calculations show that the chiral restoration and deconfinement transition 
almost coincide at low $T$ and large $\mu_B$ region \cite{Herbst2011,Xin14}. 

All these effective models describe strongly interacting
matter based on quark degrees of freedom.  Baryons are not treated in these models.
However, as far as we know, the strongly interacting matter is governed by hadronic degrees of freedom at low $T$
and small $\mu_B$. When we investigate the phase transformation
from nuclear to quark matter, it is practical to describe nuclear
matter based on the hadronic degrees of freedom at low $T$ and small $\mu_B^{}$, but
quark matter with quark-gluon degrees of freedom at high $T$ and large $\mu_B^{}$.  The phase transition boundaries can be derived
by constructing an equilibrium phase transition between hadronic and
quark matter, possibly reached in the interior of compact stars and  in HIC experiments. In the equilibrium transition, the hadronic and quark 
phases are connected through the Gibbs conditions. This approach is
widely used in the description of the phase transition in neutron
star with a quark core or kaon condensate
\cite{Glendenning91,Glendenning92, Glendenning98, Burgio02,
Maruyama07,Yang08,Shao11-1,Xu10,Dexheimer10,Shao13}. It is also generalized to explore the phase transformation from nuclear to quark matter at finite density and temperature in HICs
\cite{Muller97,Toro06,Toro09,Toro11,Liu11,Cavagnoli11,Pagliara10,Shao11-2,Shao11-3,Shao12, Shao2015}.

In our previous study
\cite{Shao11-2,Shao11-3,Shao12,Shao2015},  attention was focused on the isospin asymmetric matter, and some observable  effects on isospin-relevant meson yield
ratios were proposed. As a further study along this line, in this study we take a chemical potential dependent Polyakov loop potential in the PNJL model to construct the two-phase model and explore the full QCD phase diagram. Compared with the previous results in the original PNJL model, the calculation  presents that the phase transition lines move towards low densities (small $\mu_B$). The transition region is possibly reached in the planed experiments at the facilities of NICA/FAIR/J-PARC and BES II program at RHIC. We also analyse the transition signatures changing with the $\mu$-dependence of the Polyakov loop potential. This is another strong motivation to measure the mixed phase region in HIC experiments

The paper is organized as follows. In Sec.~II, we describe
briefly the two-phase approach and give the relevant formulas of the
Hadron-$\mu$PNJL model. In Sec.~III, we present the numerical results
about the phase diagram of the equilibrated phase transition, and
analyse the influence of the $\mu$-dependent Polyakov loop potential on the boundaries from  nuclear  to quark matter and transition signatures possibly observed in the next generation facilities .
Finally, a summary is given
in Sec. IV.

\section{ The models}
\subsection{Description of hadronic matter}
The pure hadronic matter at low $T$ and small
$\mu_B$ is described by the nonlinear Walecka type model. 
The Lagrangian is given as
\begin{eqnarray}\label{lagrangian}
\cal{L}^H_{} &\!=\!&\sum_N\bar{\psi}_N \!\big[i\gamma_{\mu}\partial^{\mu}\!- \!M
          \!+\! g_{\sigma }\sigma
                 \! -\!g_{\omega }\gamma_{\mu}\omega^{\mu}  \!-\! g_{\rho }\gamma_{\mu}\boldsymbol\tau_{}\cdot\boldsymbol
\rho^{\mu} \big]\!\psi_N       \nonumber\\
         & &    +\frac{1}{2}\left(\partial_{\mu}\sigma\partial^
{\mu}\sigma-m_{\sigma}^{2}\sigma^{2}\right)\!-\! V(\sigma)\!+\!\frac{1}{2}m^{2}_{\omega} \omega_{\mu}\omega^{\mu}
                    \nonumber\\
       & &{}
          -\frac{1}{4}\omega_{\mu\nu}\omega^{\mu\nu}  +\frac{1}{2}m^{2}_{\rho}\boldsymbol\rho_{\mu}\cdot\boldsymbol
\rho^{\mu}        -\frac{1}{4}\boldsymbol\rho_{\mu\nu}\cdot\boldsymbol\rho^{\mu\nu}   ,
 \end{eqnarray}
%\begin{widetext}
%\begin{eqnarray}\label{lagrangian}
%\cal{L}^H_{} &=&\sum_N\bar{\psi}_N[i\gamma_{\mu}\partial^{\mu}- M
  %        +g_{\sigma }\sigma
%          -g_{\omega }\gamma_{\mu}\omega^{\mu}
 %         -g_{\rho }\gamma_{\mu}\boldsymbol\tau_{}\cdot\boldsymbol
%\rho^{\mu}]\psi_N  +\frac{1}{2}\left(\partial_{\mu}\sigma\partial^
%{\mu}\sigma-m_{\sigma}^{2}\sigma^{2}\right)
%           \nonumber\\
 %      & &{}- V(\sigma)+\frac{1}{2}m^{2}_{\omega} \omega_{\mu}\omega^{\mu}
  %        -\frac{1}{4}\omega_{\mu\nu}\omega^{\mu\nu}
 %         +\frac{1}{2}m^{2}_{\rho}\boldsymbol\rho_{\mu}\cdot\boldsymbol
%\rho^{\mu}
 %         -\frac{1}{4}\boldsymbol\rho_{\mu\nu}\cdot\boldsymbol\rho^{\mu\nu}   ,
% \end{eqnarray}
%\end{widetext}
where $
\omega_{\mu\nu}= \partial_\mu \omega_\nu - \partial_\nu
\omega_\mu$, $ \rho_{\mu\nu} \equiv\partial_\mu
\boldsymbol\rho_\nu -\partial_\nu \boldsymbol\rho_\mu$. In this model, the interactions between
nucleons are mediated by $\sigma,\,\omega,\,\rho$ mesons. The self-interactions of $\sigma$ meson,
$V(\sigma)= \frac{1}{3} b\,(g_{\sigma} \sigma)^3+\frac{1}{4} c\,
(g_{\sigma} \sigma)^4$ are included to give the correct
compression modulus, the effective nucleon mass at nuclear saturation density.
The parameter set NL$\rho$ is used in the calculation, which gives a well description of  the properties of nuclear matter.
(The details can be found in Refs.~\cite{Toro09, Liu11,Toro11,Shao11-2, Shao11-3, Shao2015})

To describe asymmetric nuclear matter, we define the baryon and isospin chemical potential
as 
\begin{equation}
\mu_B^{H}=\frac{(\mu_p+\mu_n)}{2},\,\,\,\,\,\,\,\,\,  \mu_3^{H}=(\mu_p-\mu_n).
\end{equation}
The asymmetry parameter of  nuclear matter is defined
as
\begin{equation}
\alpha^{H}=(\rho_n-\rho_p)/(\rho_p+\rho_n),
\end{equation} which is determined by the heavy ions taken in experiments.
The values of  $\alpha^{H}$ are compiled for some heavy-ion sources in \cite{Cavagnoli11}, and the largest one is
 $\alpha^{H}=0.227$ in $^{238}$U+$^{238}$U collision for stable nuclei.  For unstable nuclei, $\alpha^{H}$ can take a larger value.

\subsection{Description of quark matter}
To describe pure quark matter at large $\mu_B$ and finite $T$, we use the recently developed chemical potential dependent PNJL model. 
First, we introduce the original PNJL model,
and then consider the  $\mu$-dependent Polyakov loop potential.
The Lagrangian of the standard two-flavor PNJL model is
\begin{eqnarray}\label{lagrangian-q}
\mathcal{L}^{Q}&=&\bar{q}(i\gamma^{\mu}D_{\mu}-\hat{m}_{0})q+
G\bigg[(\bar{q}q)^{2}+
(\bar{q}i\gamma_{5}\vec{\tau} q)^{2}\bigg]\nonumber \\
&&-\mathcal{U}(\Phi[A],\bar{\Phi}[A],T)
\end{eqnarray}
where $q$ denotes the quark fields with two flavors, $u$ and
$d$, and three colors; $\hat{m}_{0}=\texttt{diag}(m_{u},\ m_{d})$ in flavor space.
The covariant derivative in the Lagrangian is defined as $D_\mu=\partial_\mu-iA_\mu-i\mu_q\delta_\mu^0$.
The gluon background field $A_\mu=\delta_\mu^0A_0$ is supposed to be homogeneous
and static, with  $A_0=g\mathcal{A}_0^\alpha \frac{\lambda^\alpha}{2}$, where
$\frac{\lambda^\alpha}{2}$ is $SU(3)$ color generators.

The effective potential $\mathcal{U}(\Phi[A],\bar{\Phi}[A],T)$ is expressed in terms of the traced Polyakov loop
$\Phi=(\mathrm{Tr}_c L)/N_C$ and its conjugate
$\bar{\Phi}=(\mathrm{Tr}_c L^\dag)/N_C$. The Polyakov loop $L$  is a matrix in color space
\begin{equation}
   L(\vec{x})=\mathcal{P} \mathrm{exp}\bigg[i\int_0^{\frac{1}{T}} d\tau A_4 (\vec{x},\tau)   \bigg],
\end{equation}
where  $A_4=iA_0$.

The temperature-dependent Polyakov loop effective potential $\mathcal{U}(\Phi,\bar{\Phi},T)$  proposed in \cite{Robner07} takes the form
\begin{eqnarray}
     \frac{\mathcal{U}(\Phi,\bar{\Phi},T)}{T^4}&=&-\frac{a(T)}{2}\bar{\Phi}\Phi
                                                +b(T)\mathrm{ln}\big[1-6\bar{\Phi}\Phi\nonumber\\
                                                &&+4(\bar{\Phi}^3+\Phi^3)-3(\bar{\Phi}\Phi)^2\big],
\end{eqnarray}
where
\begin{equation}\label{T}
    a(T)=a_0+a_1\bigg(\frac{T_0}{T}\bigg)+a_2\bigg(\frac{T_0}{T}\bigg)^2,\ \  b(T)=b_3\bigg(\frac{T_0}{T}\bigg)^3.
\end{equation}
%and
%\begin{equation}
%    b(T)=b_3\bigg(\frac{T_0}{T}\bigg)^3.
%\end{equation}
The parameters $a_i$, $b_i$ summarized in Table \ref{tab:1} are
precisely fitted according to the result of lattice QCD thermodynamics in
pure gauge sector. 

\begin{table}[ht]
\tabcolsep 0pt \caption{\label{tab:1}Parameters in Polyakov effective potential given in~\cite{Robner07}}
\setlength{\tabcolsep}{10.0pt}
\begin{center}
\def\temptablewidth{0.8\textwidth}
%{\rule{\temptablewidth}{0.5pt}}
\begin{tabular}{c c c c}
\hline
\hline
   {$a_0$}                      & $a_1$        & $a_2$      & $b_3$           \\  \hline
   $ 3.51$                   & -2.47        &  15.2      & -1.75               \\ \hline
\hline
\end{tabular}
 % {\rule{\temptablewidth}{0.5pt}}
\end{center}
\end{table}
The parameter $T_0=270$\,MeV is  the confinement-deconfinement transition temperature in the pure Yang-Mills theory at  vanishing chemical potential~\cite{Fukugita90}. 
In the presence of fermions, the quantum back-reaction of the matter sector to the glue sector should be considered, which leads to a flavor and quark chemical potential dependence of the transition temperature $T_0(N_f, \mu)$  ($\mu=\mu_u=\mu_d$ for symmetric quark matter)\cite{Schaefer2007,Herbst2011,Herbst2013,Abuki08,Xin14}. By using renormalization group theory in \cite{Schaefer2007}, the form of $T_0(N_f, \mu)$ is proposed with
\begin{equation}
T_0(N_f, \mu)=T_{\tau} e^{-1/(\alpha_0 b(N_f, \mu))}
\end{equation}
where 
\begin{equation}\label{b}
b(N_f, \mu)=\frac{11N_c-2N_f}{6\pi}-\beta\frac{16N_f}{\pi}\frac{\mu^2}{T^2_\tau}.
\end{equation}
The running coupling $\alpha_0=0.304$ is fixed at the $\tau$ scale $T_\tau=$\,1.770\,GeV according to the deconfinement transition temperature $T_0=270$\, MeV of pure gauge field
with $N_f=0$ and $\mu=0$.
When fermion fields are
included, $T_0$ is rescaled to 208 MeV for 2 flavor and 187 MeV for 2+1 flavor at vanishing chemical potential. The parameter $\beta$ in Eq. (\ref{b}) governs the curvature of  $T_0(\mu)$ as a function of quark chemical potential.

With the consideration of the chemical potential dependence of Polyakov loop potential, this improved PNJL model is named the $\mu$PNJL model.
We then replace the $T_0$ with $T_0(N_f, \mu)$ in the Polyakov loop potential given in Eq. (\ref{T}).
The thermodynamical potential of quark matter in the $\mu$PNJL model within the mean field
approximation can be derived then as
%\begin{widetext}
\begin{eqnarray}
\Omega&\!=\!&\mathcal{U}(\bar{\Phi}, \Phi, T)\!+\!G({\phi_{u}\!+\!\phi_{d}})^{2}\!-\!2\int_\Lambda \frac{\mathrm{d}^{3}\boldsymbol k}
{(2\pi)^{3}}3(E_u\!+\!E_d) \nonumber \\
&&-2T \sum_{u,d}\int \frac{\mathrm{d}^{3}\boldsymbol k}{(2\pi)^{3}} \bigg[\mathrm{ln}(1+3\Phi e^{-(E_i-\mu_i)/T}  \nonumber \\
&&+3\bar
{\Phi} e^{-2(E_i-\mu_i)/T}+e^{-3(E_i-\mu_i)/T}) \bigg]\nonumber \\
&&-2T \sum_{u,d}\int \frac{\mathrm{d}^{3}\boldsymbol k}{(2\pi)^{3}} \bigg[\mathrm{ln}(1+3\bar{\Phi}
e^{-(E_i+\mu_i)/T} \nonumber \\
&& +3\Phi e^{-2(E_i+\mu_i)/T}+e^{-3(E_i+\mu_i)/T}) \bigg],
\end{eqnarray}
%\end{widetext}
%\begin{widetext}
%\begin{eqnarray}
%\Omega&=&\mathcal{U}(\bar{\Phi}, \Phi, T)+G(\Phi)({\phi_{u}+\phi_{d}})^{2}-2\int_\Lambda \frac{\mathrm{d}^{3}\boldsymbol k}
%{(2\pi)^{3}}3(E_u+E_d) \nonumber \\
%&&-2T \sum_{u,d}\int \frac{\mathrm{d}^{3}\boldsymbol k}{(2\pi)^{3}} \bigg[\mathrm{ln}(1+3\Phi e^{-(E_i-\mu_i)/T}+3\bar
%{\Phi} e^{-2(E_i-\mu_i)/T}+e^{-3(E_i-\mu_i)/T}) \bigg]\nonumber \\
%&&-2T \sum_{u,d}\int \frac{\mathrm{d}^{3}\boldsymbol k}{(2\pi)^{3}} \bigg[\mathrm{ln}(1+3\bar{\Phi}
%e^{-(E_i+\mu_i)/T}+3\Phi e^{-2(E_i+\mu_i)/T}+e^{-3(E_i+\mu_i)/T}) \bigg],
%\end{eqnarray}
%\end{widetext}
where $E_i=\sqrt{\boldsymbol k^{\,2}+M_i^2}$ is energy-momentum dispersion relation
of quark flavor $i$, and $\mu_i$ is the corresponding quark chemical
potential.

The dynamical quark masses and quark condensates are
coupled with the following equations
\begin{equation}\label{mass}
M_i=m_{0}-2G(\phi_u+\phi_d),
\end{equation}
\begin{equation}
\phi_i=-2N_{c}\int\frac{d^{3}\boldsymbol k}{(2\pi)^{3}}\frac{M_i}{E_i}
\big(1-n_i(k)-\bar{n}_i(k)\big),
\end{equation}
where $n_i(k)$ and $\bar{n}_i(k)$ 
%\begin{widetext}
\begin{equation}\label{distribution}
  n_{i}(k)\!=\!\frac{\Phi e^{\!-\!(E_i\!-\!\mu_i)/T}\!+\!2\bar{\Phi} e^{\!-\!2(E_i\!-\!\mu_i)/T}+e^{-3(E_i-\mu_i)/T}}
  {1\!+\!3\Phi e^{\!-\!(E_i\!-\!\mu_i)/T}\!+\!3\bar{\Phi} e^{\!-\!2(E_i\!-\!\mu_i)/T}\!+\!e^{\!-\!3(E_i\!-\!\mu_i)/T}} , 
\end{equation}
\begin{equation}
  \bar{n}_{i}(k)\!=\!\frac{\bar{\Phi} e^{\!-\!(E_i\!+\!\mu_i)/T}\!+\!2{\Phi} e^{\!-\!2(E_i\!+\!\mu_i)/T}\!+\!e^
  {-3(E_i+\mu_i)/T}}{1\!+\!3\bar{\Phi} e^{\!-\!(E_i\!+\!\mu_i)/T}\!+\!3{\Phi} e^{\!-\!2(E_i\!+\!\mu_i)/T}\!\!+\!\!e^{\!-\!3(E_i\!+\!\mu_i)/T}} .
\end{equation}
%\end{widetext}
are modified Fermion distribution functions of
quark and antiquark. The values of $\phi_u, \phi_d, \Phi$ and $\bar{\Phi}$ can be determined
by minimizing the thermodynamical potential
\begin{equation}
\frac{\partial\Omega}{\partial\phi_u}=\frac{\partial\Omega}{\partial\phi_d}=\frac{\partial\Omega}
{\partial\Phi}=\frac{\partial\Omega}{\partial\bar\Phi}=0.
\end{equation}
All the thermodynamic quantities relevant to the bulk properties of
quark matter can be obtained from $\Omega$. Particularly, we note that the pressure
and energy density should be zero in the vacuum.
In the calculation a cut-off $\Lambda$ is implemented in 3-momentum
space for divergent integrations.
$\Lambda=651$ MeV, $G=5.04\,\mathrm{GeV}^{-2}$,
$m_{u,d}=5.5$\,MeV will be taken by fitting the experimental values of pion decay constant $f_{\pi}=92.3$\,MeV and pion mass $\ m_{\pi}=139.3$\,MeV \cite{Ratti06}.

For asymmetric quark matter,  the baryon and isospin chemical potential are defined as
 $\mu_B^{Q}=\frac{3}{2}(\mu_u+\mu_d)$, $\mu_3^{Q}=(\mu_u-\mu_d)$, respectively. The quark chemical potential $\mu$ in $T_0(N_f, \mu)$  can take the mean values of $u,\,d$ quark. The asymmetry parameter of pure quark matter is
\begin{equation}
\alpha^{Q}=-\frac{\rho_{3}^{Q}}{\rho_{B}^{Q}}=-\frac{(\rho_u-\rho_d)}{(\rho_u+\rho_d)/3}
\end{equation}
 where $\rho_{3}^{Q}=(\rho_u-\rho_d)$, and $\rho_{B}^{Q}=(\rho_u+\rho_d)/3$.

\subsection{Transformation from hadronic to quark matter}
The above is a separate description of the purely hadronic  and  quark
matter. When the equilibrium transition between the hadronic and quark matter forms,
the Gibbs' conditions with the thermal, chemical and mechanical equilibrium need to be satisfied (A general discussion
of phase transitions in multicomponent systems can be found in Ref.~\cite{Glendenning92}),
\begin{equation}\label{Gibbs}
\mu_B^H=\mu_B^Q,\  \ \mu_3^H=\mu_3^Q,\  \ T^H=T^Q,\ \ P^H=P^Q,
\end{equation}
where $\mu_3^H$ and $\mu_3^Q$ are the isospin chemical potential of the hadronic and quark phase, separately.
In the coexisting region, the total baryon density is consisted of two parts, $\rho_B^{}=(1-\chi)\rho_B^{H}+\chi \rho_B^{Q}$ where $\chi$
is the fraction of quark matter, and $1-\chi$ is the  ratio of nuclear matter. Similarly,
$\rho_3^{}=(1-\chi)\rho_3^{H}+\chi \rho_3^{Q}$
is the total isospin density.
%\begin{eqnarray}\label{tcm}
%& &\mu_B^H(\rho_B^{},\rho_3^{},T)=\mu_B^Q(\rho_B^{},\rho_3^{},T)\nonumber\\
%& &\mu_3^H(\rho_B^{},\rho_3^{},T)=\mu_3^Q(\rho_B^{},\rho_3^{},T)\nonumber\\
%& &P^H(\rho_B^{},\rho_3^{},T)=P^Q(\rho_B^{},\rho_3^{},T)?
%\end{eqnarray}
%where the symbles of $H$ and $Q$ means hadron and quark.

As shown in the previous study \cite{Muller97,Toro06,Toro09,Toro11,Liu11,Cavagnoli11,Pagliara10,Shao11-2,Shao11-3,Shao12, Shao2015}, the phase transition features of asymmetric matter are isospin dependent.
Once the species of heavy ions are chosen in HIC experiments, the asymmetry parameter will be determined.
Due to the isospin conservation in strong interaction, the global asymmetry parameter $\alpha$ 
\begin{eqnarray}\label{isospin}
& &   \alpha\equiv-\frac{\rho_{3}^{}}{\rho_{B}^{}}= -\frac{(1-\chi)\rho_3^{H}+
\chi \rho_3^{Q}}{(1-\chi)\rho_B^{H}+\chi \rho_B^{Q}},
\end{eqnarray}
for the mixed phase should maintain constant.
However, in the coexisting region the local
asymmetry parameters, $\alpha^H$ and $\alpha^Q$, can vary for
different quark fraction  $\chi$. It's just the $\chi$-dependence of $\alpha^H$ and $\alpha^Q$ that provides the
possibility to test the isospin relevant signals generated in the
hadronization stage in HIC experiments. The details about the phase
transformation from asymmetric nuclear matter to quark matter will
be discussed in the next section. One can also refer to our previous
researches \cite{Toro11,Shao11-2, Shao11-3, Shao12, Shao2015}

\section{Numerical results and discussions }
\subsection{Features of pure quark matter in the $\mu$PNJL model}
\begin{figure}[htbp]
\begin{center}
\includegraphics[scale=0.29]{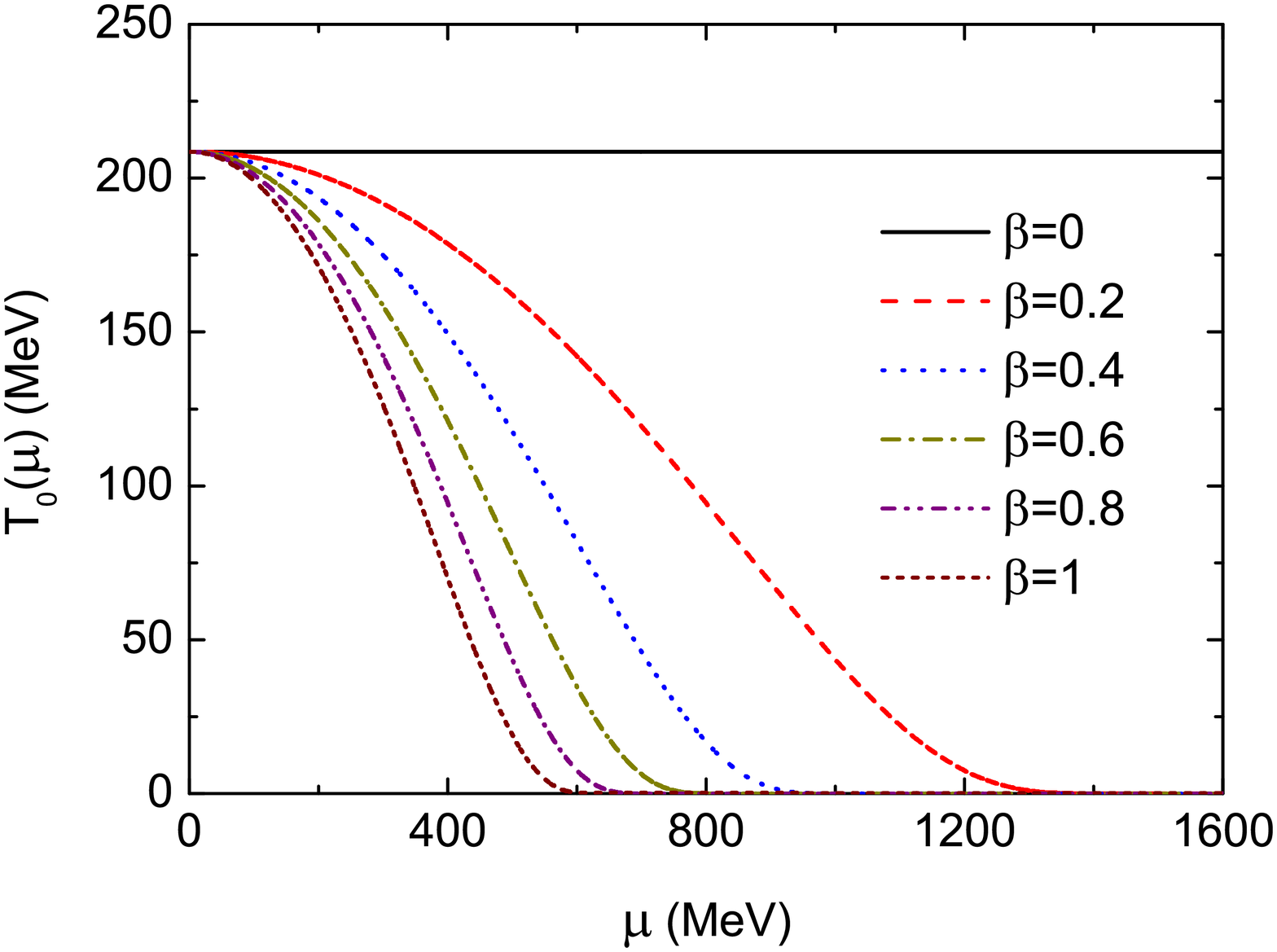}
\caption{\label{fig:T0-mu}(color online) $T_{0}(\mu)$
as a function of $\mu$ with various $\beta$ from 0 to 1. The case $\beta=0$ corresponds to the standard PNJL model.}
\end{center}
\end{figure}

In this subsection, we present some properties of pure symmetric
quark matter in the $\mu$PNJL model. First, we present in Fig.~\ref{fig:T0-mu} the chemical potential dependence of $T_0(\mu)$. 
Different values of $\beta$ are used in the calculation for a tentative study. In the case of $\beta=0$, corresponding to the standard PNJL model in which only the contribution from gauge field to Polyakov loop potential is considered, $T_0(\mu)=208$\,MeV is a constant, as shown with the solid line in Fig. \ref{fig:T0-mu}. The dotted lines show the results for $\beta \neq 0$.  This figure manifests that  $T_0(\mu)$ is sensitive to $\beta$ which in some degree can be taken as a parameter to reflect the interaction strength between matter sector and glue sector.

\begin{figure}[htbp]
\begin{center}
\includegraphics[scale=0.28]{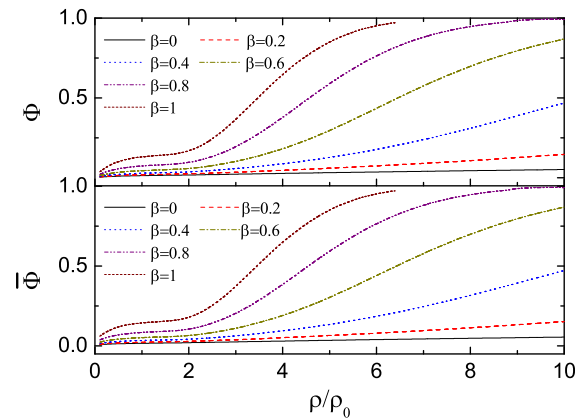}
\caption{\label{fig:phi-rho}(color online) Polyakov loop 
$\Phi$ and $\bar{\Phi}$
as functions of baryon density $\rho_B$ for different $\beta$ at $T$=20 MeV.}
\end{center}
\end{figure}

Fig.~\ref{fig:phi-rho} presents the values of Polyakov loop $\Phi$ and $\bar{\Phi}$ as functions of baryon density for various $\beta$  at $T=$20\,MeV. 
In the original PNJL model (the case $\beta=0$), $\Phi$ and $\bar{\Phi}$ always take small values at low temperature. This means that quarks are confined, even in the high density region where the chiral symmetry is restored already. This forms the so-called quarkyonic phase at low $T$ and high density region. However with the consideration of quark back-reaction to glue sector, quark confinement-deconfinement phase transition can occur at low $T$, as shown by the dotted lines with different values of $\beta$. If we take the standard that $\Phi$ or $\bar{\Phi}$=0.5 marks the happening of deconfinement transition, as adopted in \cite{Fukushima04, Sakai10}, we find the transition density moves to  a lower one for a larger $\beta$. If we take $\beta\gg1$ in the calculation, unphysical results will be derived with too small deconfined baryon density where the chiral symmetry is still breaking. For more details about the properties of quark matter, one can refer to \cite{Abuki08,Xin14}. In this study we mainly emphasize the transformation from hadronic to quark matter at intermediate densities  in the two-phase model.

\subsection{Transition boundaries from hadronic to quark matter in the two-phase model with different  $
\beta$}

In this part we focus on the phase transition from asymmetric nuclear to quark matter in the two-phase model. Since the largest
asymmetry parameter could be reached for stable nuclei is  $\alpha = 0.227$ in $^{238}$U+$^{238}$U collision,  we choose $\alpha = 0.2$ and different $\beta$ to demonstrate the features of the phase transition of asymmetric matter. As a matter of fact, $\alpha$ can take
a larger value for neutron-rich unstable nuclei.

\begin{figure}[htbp]
\begin{center}
\includegraphics[scale=0.28]{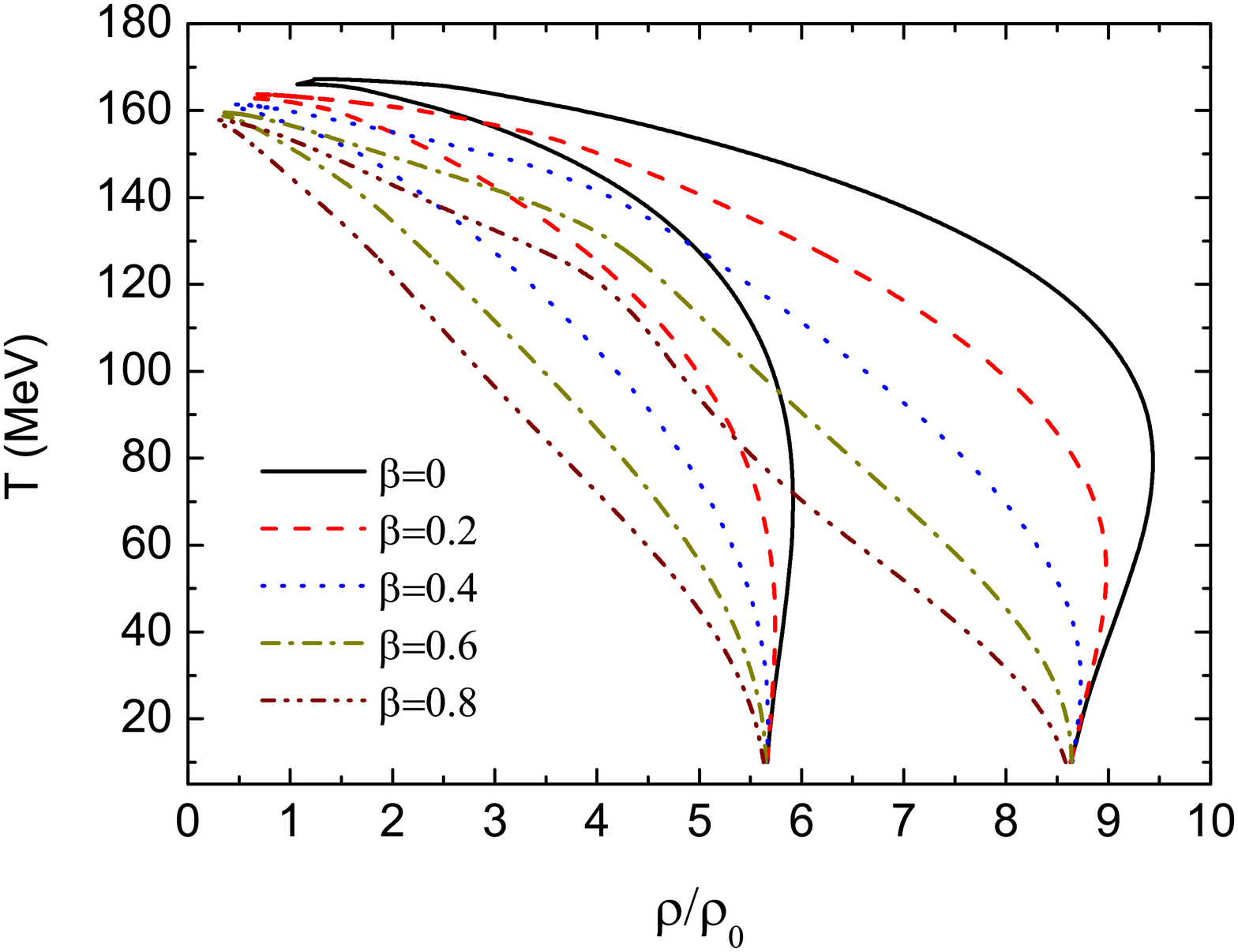}
\caption{\label{fig:T-rho-beta}(color online) Phase diagram of the hadron-quark phase transition in the T-$\rho_B$ plane for different $\beta$ with the asymmetry parameter $\alpha$=0.2.}
\end{center}
\end{figure}

\begin{figure}[htbp]
\begin{center}
\includegraphics[scale=0.28]{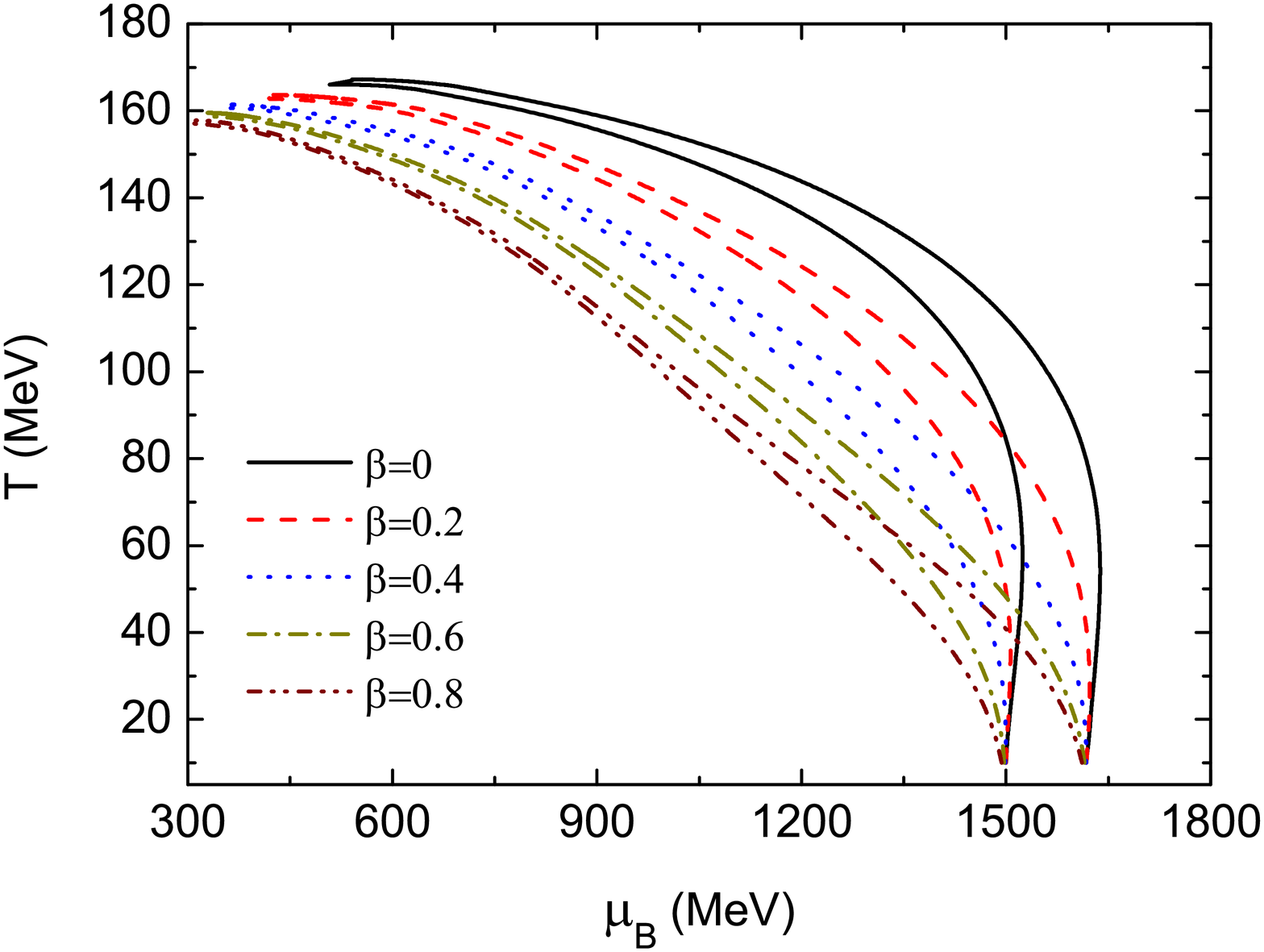}
\caption{\label{fig:T-mu-beta}(color online) Phase diagram of the hadron-quark phase transition in the T-$\mu_B$ plane for different $\beta$ with the asymmetry parameter $\alpha$=0.2.}
\end{center}
\end{figure}

The equilibrium phase transition is constructed based on  Gibbs criteria given in Eq.~(\ref{Gibbs}) and the isospin charge conservation given in Eq.~(\ref{isospin}) for strong interaction.
Fig.~\ref{fig:T-rho-beta} and Fig.~\ref{fig:T-mu-beta} show the  boundaries of hadron-quark transition in the $T-\rho_B$ and $T-\mu_B$  diagram with a series of $\beta$. For each value of $\beta$, the curves with the same color mark the boundaries of purely hadronic matter (at low densities )  and purely quark matter (at high densities).
For the equilibrium transition derived with the original PNJL model (the case $\beta=0$), the transition lines as functions of $\rho_B$ and $\mu_B$ vary non-monotonously with the increase of $T$. This feature maintains when a weak interaction between matter sector and glue sector is included, e.g., in the case of $\beta=0.2 $ and $ 0.4$. But the transition lines as functions of $\rho_B$ and $\mu_B$ decrease monotonously when $\beta\geqslant$0.6 is taken. These features indicate that the back-reaction of matter sector to glue sector is crucial and indispensable  for the hadron-quark phase transition.
 
In Fig.~\ref{fig:T-rho-beta} and Fig.~\ref{fig:T-mu-beta}, the region between the curves with the same color  for each  $\beta$ is the hadron-quark coexisting phase. The transition in the coexisting phase is the first order because of the discontinuity of baryon density in the two phases. Fig.~\ref{fig:T-rho-beta} and Fig.~\ref{fig:T-mu-beta}  also present that, with the increase of $\beta$, the transition boundaries move towards smaller $\rho_B$ and $\mu_B$. In particular,  the end point of the phase transition  moves also towards lower $T$  with the increase of $\beta$. The main reason is that the confinement-deconfinement transition temperature $T_0(\mu)$ decreases to a lower value when a larger parameter of $\beta$ is taken, as shown in Fig.~\ref{fig:T0-mu}. 

\begin{figure}[htbp]
\begin{center}
\includegraphics[scale=0.27]{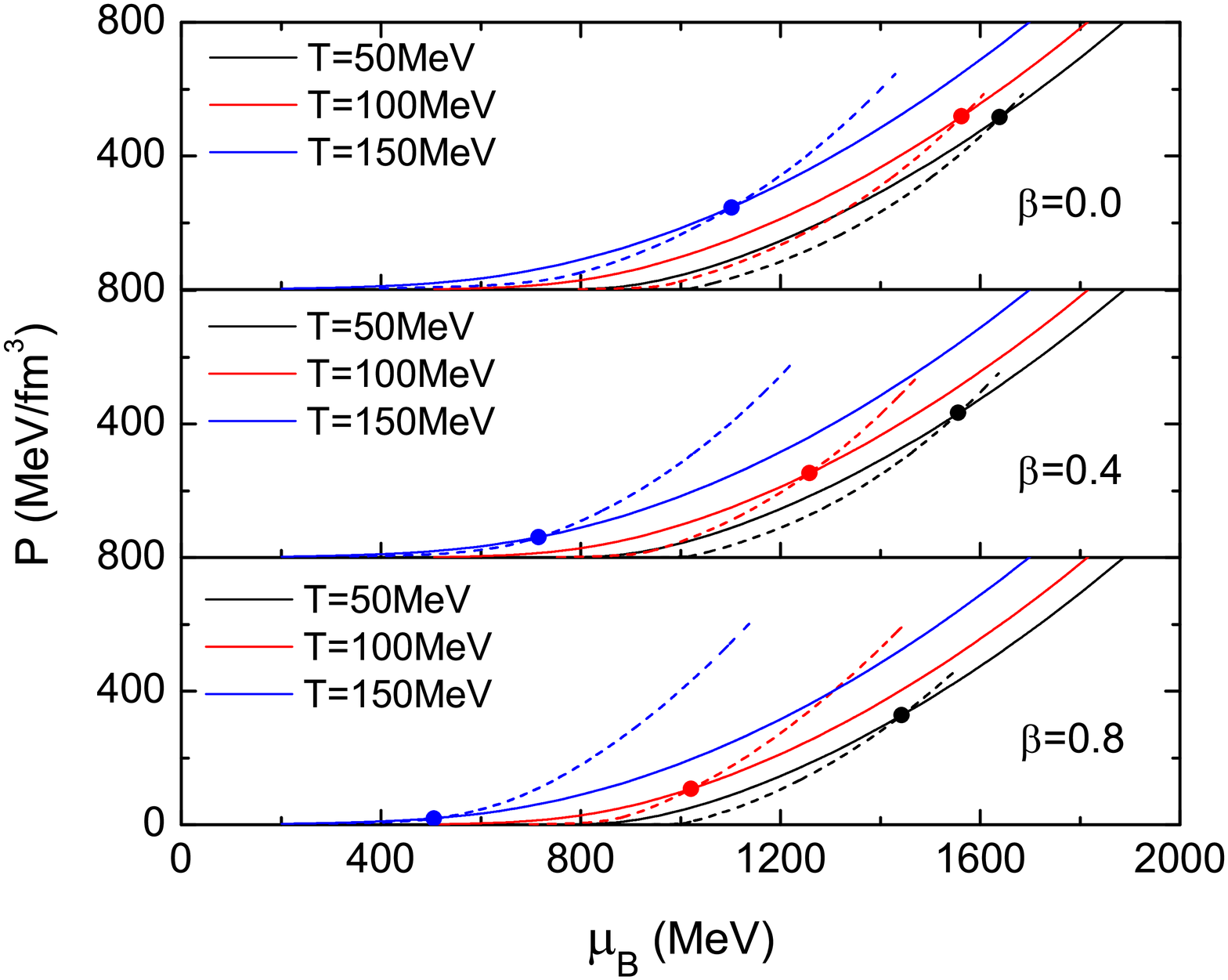}
\caption{\label{fig:P-mu}(color online) Pressure of symmetric hadronic (solid) and quark pressure (dashed) as functions of baryon chemical potential at T=50 MeV, 100 MeV and 150 MeV  for several $\beta$.}
\end{center}
\end{figure}

 To further understand the effect of $\beta$ on the phase transition, we plot the $P-\mu_B$ phase diagram of symmetric nuclear and quark matter in Fig.~\ref{fig:P-mu}. 
 The solid and dashed curves are the pressure of pure hadronic and quark matter, respectively.  The solid dots indicate the locations of the hadron-quark phase transition where the Gibbs criteria are fulfilled at different conditions. For a given temperature, we can see that the phase transition (solid dot) moves to a smaller $\mu_B$ with the increase of $\beta$. The reason is the pressure of quark matter increases more quickly when a larger $\beta$ is taken, and then the mechanical equilibrium can be reached at a smaller $\mu_B$.  On the other hand, for a given $\beta$, with the increase of temperature the phase transition also moves towards a smaller $\mu_B$, since the increase of the pressure of quark matter is faster than that of hadronic matter. The intersection point (solid dot) will finally vanish when the temperature is higher than a critical value beyond which the pressure of quark phase will be always higher than that of hadronic phase, then the phase equilibrium cannot be reached any more. Therefore, there is an end point and the mixed phase finally disappears at high temperature in the two-phase model.

Similar results can be obtained for asymmetric matter. In addition, we note that there is a critical value of $\beta=0.89$.  For the case $\beta>0.89$, the equilibrium transition cannot be realized for $\alpha=0.2$ in the two-phase model. The measurement of the mixed phase  in HIC experiments possibly provide relevant information on the $\beta$ parameter, i.e. on the $\mu$-dependence of the Polyakov loop potential.
 
Furthermore, we give a comparison of the phase transition in the $\mu$PNJL model and the two-phase model.
The confinement-deconfinement phase transition can be realized at low $T$ in the $\mu$PNJL model, but the threshold depends on the parameter $\beta$, which reflects the interaction strength of matter sector to glue sector. Compared with the two-phase model, if a smaller $\beta$ is taken,  the onset density of deconfinement phase transition in the  $\mu$PNJL model  will be larger than that in the two-phase model. But the phase transition line moves towards lower densities with the increase of $\beta$. The onset density can be even smaller than that of the two-phase model if $\beta$ is large enough.

For the case $\beta=0.4$ or smaller, when the hadron-quark phase transition happens in the two-phase model as shown in Fig.~\ref{fig:T-rho-beta}, the value of $\Phi$ and $ \bar{\Phi}$ in the quark phase are still smaller  than 0.5 as shown in Fig.~\ref{fig:phi-rho}, it seems that the two-phase model predicts a transition to confined quark-matter. The appearance of such a behavior is attributed to the quark model. We know that,  in the original PNJL model the confinement-deconfinement phase transition cannot be realized at very low $T$ because   $\Phi$ and $ \bar{\Phi}$ always take small values. This is one of the most important reasons that motivate us take the $\mu$PNJL model. However, even in the $\mu$PNJL model with $\beta=0.4$ or smaller, the confinement-deconfinement phase transition at low $T$ can only happen at very high baryon number densities. This explains why it seems that the two-phase model predicts a transition to confined quark-matter for a small $\beta$ in the two-phase model.  In principle, the two-phase model  describes the phase equilibrium where quarks have deconfined.
From this point of view, in the two-phase model the results derived with   $\beta=0.4$ or smaller is unphysical. Therefore, it is necessary to introduce the $\mu$-dependent $T_0$ and take a value of $\beta$ larger than 0.4, which also means the back-reaction of matter to glue sector is strong at finite densities .

We also note that the two kinds of phase transitions are constructed in different methods. The two-phase model describes the equilibrium phase transition possibly reached during the formation of quark matter in heavy-ion collisions. The deconfinement phase transition in the PNJL or $\mu$PNJL quark model is derived based on the quark-gluon degrees of freedom. To what degree the phase transition from deconfinement to confinement in the quark model can be identified with the formation of hadronic matter is still not clear.  For example, there exists the so-called  ``quarkyonic phase'' in the PNJL model where  the chiral symmetry has restored but quarks are still confined at low $T$, and the ``coincidence problem'' exists at high $T$.  We emphasize that related discussions are still open issues. The experiments in the future will provide us more information about  the phase transition.

\subsection{Effects of Isospin asymmetry}

Now  we discuss the influence of asymmetry parameter $\alpha$ on the
phase diagram. Different values of $\alpha$ correspond to different kinds of heavy-ion sources  chosen in HIC experiments. Considering the unstable nuclei, $\alpha$ can take a value larger than 0.227. Therefore, we  take the values of $\alpha$ between 0 and 0.35 in the calculation to show the isospin effect.
\begin{figure}[htbp]
\begin{center}
\includegraphics[scale=0.28]{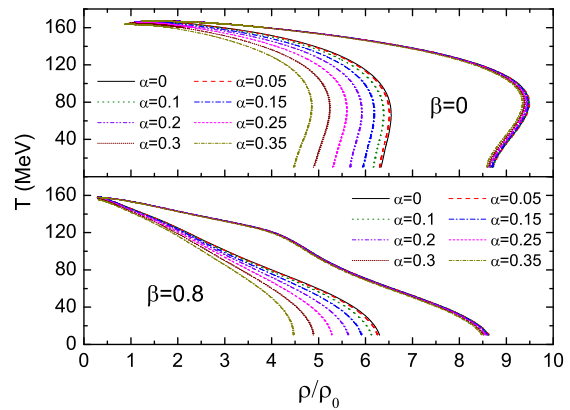}
\caption{\label{fig:2D-T-rho}(color online) Phase diagrams of the hadron-quark phase transition in the T-$\rho_B$ plane with various asymmetry parameters for $\beta$=0  (upper panel) and $\beta$=0.8 (lower panel).}
\end{center}
\end{figure}

\begin{figure}[htbp]
\begin{center}
\includegraphics[scale=0.28]{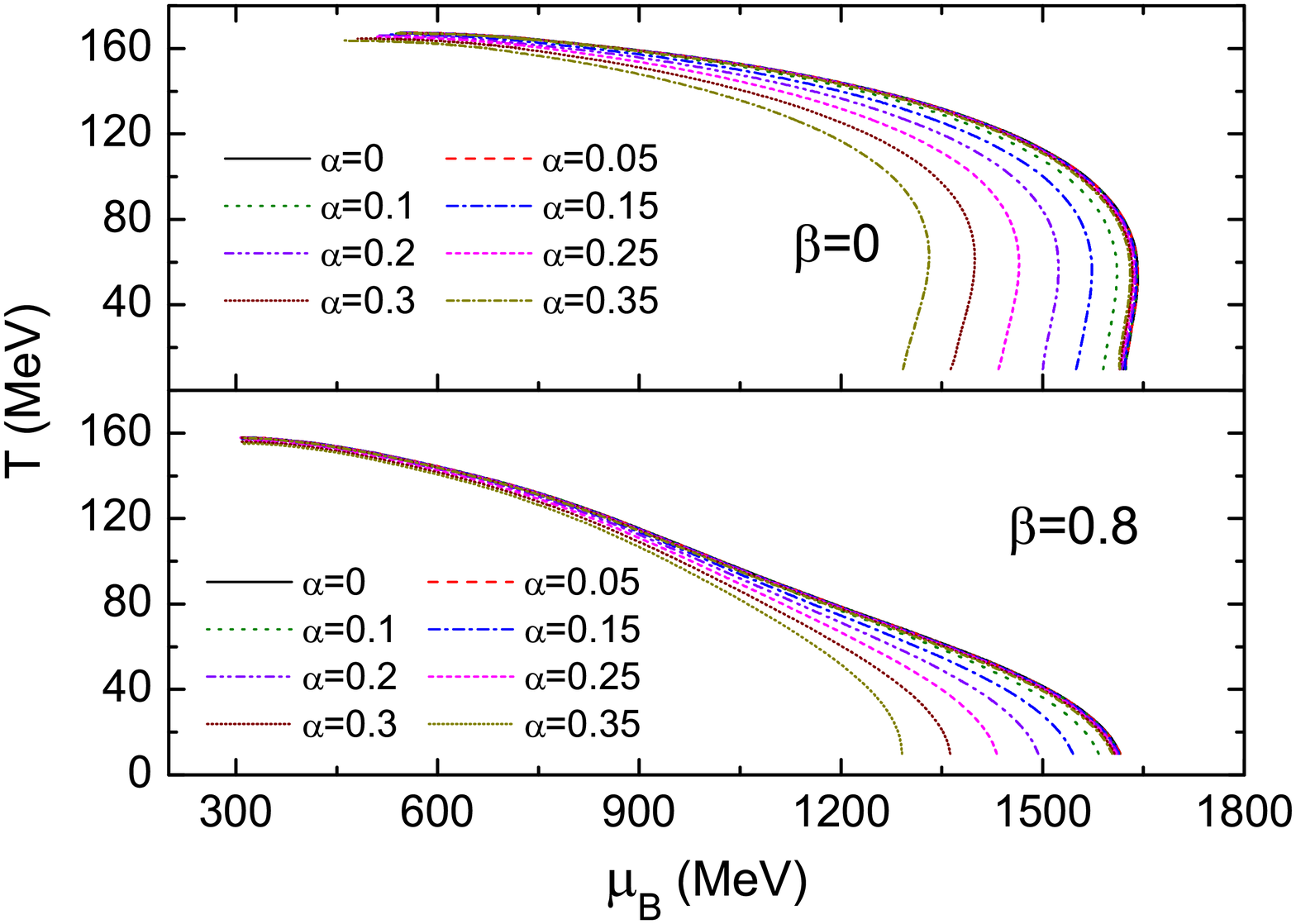}
\caption{\label{fig:2D-T-mu}(color online) Phase diagrams of the hadron-quark phase transition in the T-$\mu_B$ plane with various asymmetry parameters for $\beta$=0 (upper panel) and $\beta$=0.8 (lower panel).}
\end{center}
\end{figure}

We plot the phase diagram of equilibrium transition from asymmetric hadronic to quark matter in Fig.~\ref{fig:2D-T-rho} and Fig.~\ref{fig:2D-T-mu} in the two-phase model.
In the two figures, the upper panels are the results of the original PNJL model. The lower panels are the results of the $\mu$-dependent PNJL model with the contribution of both matter sector and glue sector with $\beta=0.8$. In addition to the conclusion derived in the last subsection, the two figures show that the onset densities (chemical potentials) move to smaller ones with the increase of asymmetry parameter $\alpha$, and the corresponding coexisting region enlarges. On the other hand, for each value of $\alpha$ the coexisting region shrinks greatly at high temperature. We note that  for symmetric matter (the case $\alpha=0$), only one transition line but not transition region exists in the $T-\mu_B$ diagram.

\begin{figure}[htbp]
\begin{center}
\includegraphics[scale=0.27]{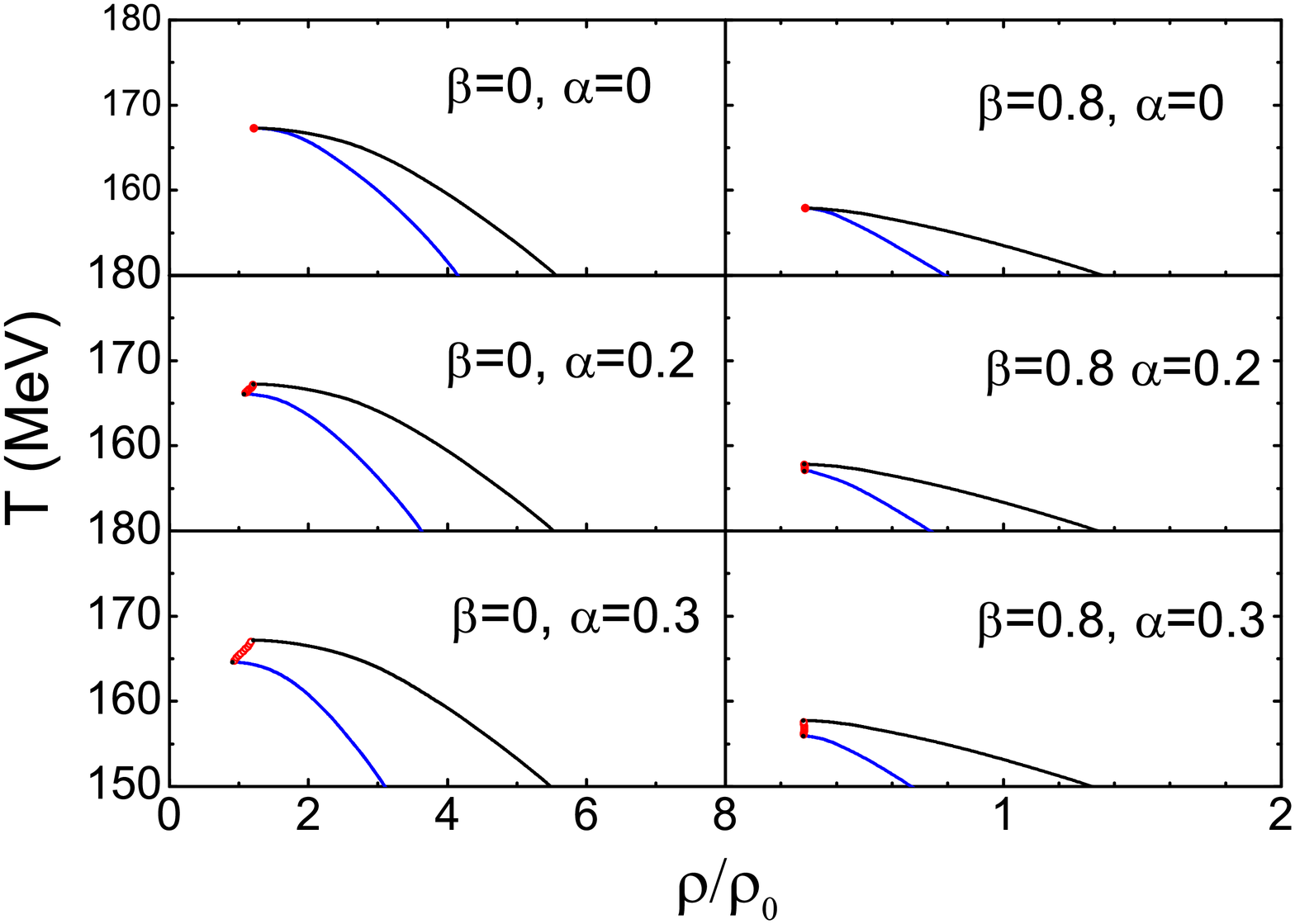}
\caption{\label{fig:T-rho-high-T}(color online) Critical behavior of the hadron-quark phase transition for symmetric and asymmetric matter with different $\alpha$, $\beta$.}
\end{center}
\end{figure}

To show more clearly how the asymmetry parameter $\alpha$ affect the phase transition  near the end point, we present in Fig.~\ref{fig:T-rho-high-T} the details of $T-\rho_B$ phase diagram in the high-temperature region. The upper panel of Fig.~\ref{fig:T-rho-high-T} are the results of symmetric matter with $\alpha=0$, and the middle and lower panel are the results of asymmetric matter with $\alpha=0.2$ and 0.3, respectively. The upper panel shows that the boundaries of $\chi=0$  and $\chi=1$ have the same end point at high temperature for symmetric matter. However, for asymmetric matter, the locations of the end points for $\chi=0$  and $\chi=1$ are slightly different for asymmetric matter,  as shown with the solid dots in the middle panel for $\alpha=0.2$ and in the lower panel for $\alpha=0.3$. In Fig.~\ref{fig:2D-T-rho} and Fig.~\ref{fig:2D-T-mu}  (including also the relevant phase diagram in our previous study  \cite{Shao11-2,Shao11-3,Shao12,Shao2015}, ) only the boundaries with $\chi=0$ and $\chi=1$ are plotted. As matter of fact, we can find an end point for each value of $\chi$ in the region $0<\chi<1$. For a given $\chi$ in the calculation, before the end point is reached, $\rho_B^H < \rho_B^Q$ can be derived for the phase equilibrium in the mixed phase.
The two curves of $T-\rho_B^H$ and $T-\rho_B^Q$ intersect at one point (the end point) at high T.  Therefore, there exists an critical end point for each $\chi$ for asymmetric matter. It means that the end point is $\chi$ dependent for asymmetric matter. If these end points for different $\chi$ are connected, a short phase transition line forms as shown with the red bubbles in the middle and lower panels in Fig.~\ref{fig:T-rho-high-T}. 

\subsection{Isospin distillation effect in the mixed phase and observables in the hadronization}

Now we discuss the isospin distillation effect in the hadron-quark coexisting phase of asymmetric matter. It is related to the isospin-relevant observables in the hadronization process.
In the two-phase model, the asymmetry parameter $\alpha$ is globally conserved in the mixed phase, but the local asymmetry parameter $\alpha^{H}$ and $\alpha^{Q}$ can vary with the changing of quark fraction $\chi$ during the phase transition. We fix $\alpha=0.2$ in the following calculation to explore the isospin distillation effect with different $\beta$.
We note that the results for $\beta>0.4$ is required with the presupposition that  the phase transition is from nuclear matter to deconfined quark matter. 

\begin{figure}[htbp]
\begin{center}
\includegraphics[scale=0.3]{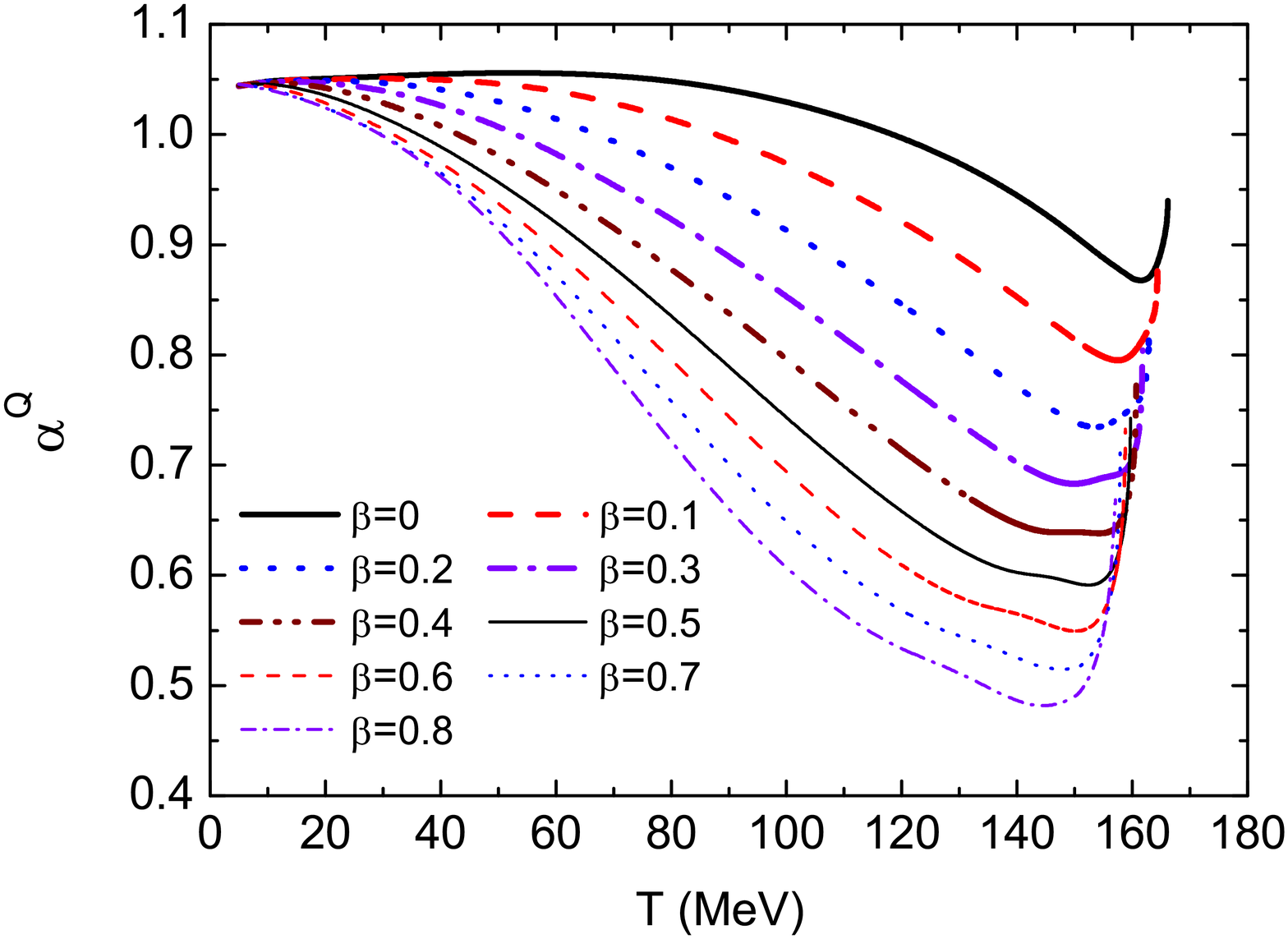}
\caption{\label{fig:alfa-T}(color online) Local isospin asymmetry parameter $\alpha^{Q}$ in the mixed phase with quark fraction $\chi$=0.01 as a function of temperature for various $\beta$ with the global asymmetry parameter $\alpha$=0.2.}
\end{center}
\end{figure}

Fig.~\ref{fig:alfa-T} shows the local asymmetry parameter $\alpha^{Q}$ as a function of $T$  at the beginning of the transition with $\chi=0.01$ for different values of $\beta$. This figure demonstrates that $\alpha^{Q}$ decreases with the increase of $\beta$. This behavior is relevant to the symmetry energy of both nuclear and quark matter. Our previous calculation~\cite{Shao11-2} shows that the symmetry energy of nuclear matter increases quickly with the rising baryon density, but that of quark matter increases slightly.  When a larger $\beta$ is taken to construct the equilibrium transition, the phase equilibrium moves to smaller $\rho_B$ and $\mu_B$ as shown in Fig.~\ref{fig:T-rho-beta} and Fig.~\ref{fig:T-mu-beta}. Then the difference of the symmetry energy between nuclear and quark matter at low density (small $\mu_B$) is relatively smaller, which result in the reduction of $\alpha^{Q}$. For more details, one can refer to~\cite{Shao11-2}.

\begin{figure}[htbp]
\begin{center}
\includegraphics[scale=0.31]{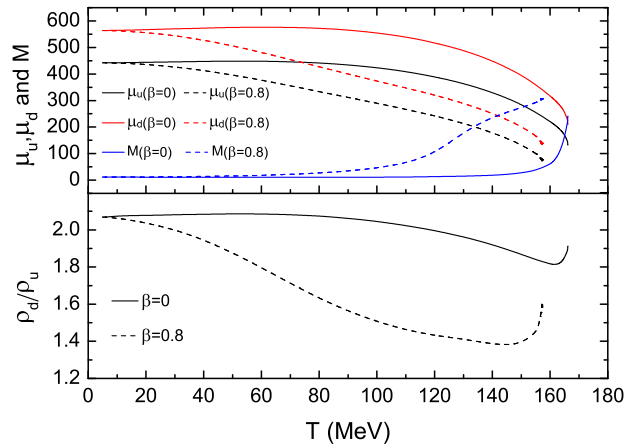}
\caption{\label{fig:muandmas-T}(color online) (upper panel) $u, d$ quark chemical potentials and dynamical quark mass as well as $\rho_d/\rho_u$ (lower panel) as functions of $T$ inside the mixed phase at $\chi$=0.01 with $\alpha$=0.2 for $\beta$=0 and $\beta$=0.8, respectively.}
\end{center}
\end{figure}

Fig.~\ref{fig:alfa-T} also presents that for each $\beta$ there is an inflection point of $\alpha^{Q}$  in the region of high $T$ (corresponding to low $\rho_B$ and small $\mu_B$). This behavior reflects the relations between dynamical quark mass and $u\,,d$ quark chemical potential. 
To see more clearly this phenomenon, we plot in Fig.~\ref{fig:muandmas-T} the evolution curves of the dynamical quark mass and $u\,,d$ quark chemical potential~(upper panel)  as well as the ratio of  $\rho_d/\rho_u$~(lower panel),  as functions of temperature in the mixed phase with a fixed $\chi=0.01$. Fig.~\ref{fig:muandmas-T} shows that the dynamical  quark mass in the equilibrium transition increases with the rising $T$ (towards low $\rho_B$ and small $\mu_B$). It implies that the chiral dynamics is affected by both temperature and density (chemical potential). %One can refer to \cite{Shao2011c} for the contour of quark dynamical mass in the $T-\rho_B$ diagram. 
From Fig.~\ref{fig:muandmas-T} we can also see that both $u$ and $d$ quark chemical potentials decrease with the rising $T$. Because $\mu_u$ is  smaller than $\mu_d$,  the intersection of the curves of $\mu_u$ and quark mass $M$ appear at a relatively lower $T$ than that of $\mu_d$ and $M$. If the system is a simple fermion system with the fermion distribution function $f=1/(1+e^{(E-\mu)/T})$,  $\rho_u$ will sharply decrease when $\mu_u$  is smaller than the dynamical quark mass $M$ at low $T$. Correspondingly, an inflection point of $\alpha^{Q}$~(with the difinition of $\alpha^{Q}=(\rho_d-\rho_u)/(\rho_u+\rho_d)/3$) at the intersection point of  $\mu_u$ and $M$ should appear.  However, such a behavior is not seen in Fig.~\ref{fig:muandmas-T}. As a matter of fact, for $\beta$=0.8 (0), the fast increase  of $\rho_d/\rho_u$ occurs for temperatures larger (smaller) than the one
corresponding to the crossing point of $\mu_u$ and $M$.  The deviation of the inflection point from $\mu_u=M$  shows that the quark system can not be take as a simple fermion system. There exists complex interactions as indicated by  the quark distribution function given in Eq.~(\ref{distribution}).  Besides $\mu$ and $M$,  Eq.~(\ref{distribution}) indicates that the quark distribution function is also relevant to $\Phi$ and $\bar{\Phi}$ which are affected by the temperature and quark chemical potential
(i.e., by the parameter $\beta$).  On the other hand, since the temperature at the inflection point is in the region about $(150-165)$\,MeV, the thermal excitation to some degree alleviates the fast decrease  of the quark distribution function at the crossing point of $\mu_u=M$. Therefore, the location of the inflection point is much more complex than that in a simple fermion system.

If the phase equilibrium can be reached in HIC experiments, the generated quarks will recombine into hadrons in the later hadronization process, and a sudden increase of isospin relevant meson yield ratio such as $\pi^- /
\pi^+$, $K^0/K^+$ possibly be observed since the isospin enrichment in the quark phase is well present (in particular in the initial part of the mixed phase). The strength of these signals depends on the value of $\alpha^Q$ (or $\rho_d/\rho_u$) which is relevant to the parameter $\beta$ in the $\mu$PNJL model. Compared with the original PNJL model, the introduction of $\mu$-dependent Polyakov loop potential (with $\beta>0.4$) will weaken the isospin relevant signals at high $T$, leading to a relatively smaller ratio of $\pi^- /
\pi^+$, $K^0/K^+$.  However, the isospin effect is still distinct at low $T$ as demonstrated in Fig.~\ref{fig:alfa-T}
and  the lower panel of  Fig.~\ref{fig:muandmas-T}. 
There is also an inflection point of these isospin relevant signals at high temperature as indicated in Fig. 9 and 10. For more discussion about the phase transition signatures in the two-phase model, one can refer to \cite{Toro16} prepared for the NICA whitepaper.

Due to the isospin conservation during the phase transition with strong interaction,  the global isospin asymmetry parameter $\alpha$ given by Eq.~(\ref{isospin}) for the mixed phase  
includes the contribution from both nuclear and quark component. For a given heavy ion source in HIC experiments, i.e., a given $\alpha$, if the local asymmetry parameter
$\alpha^Q$ is larger when the equilibrium is reached, the local asymmetry parameter $\alpha^H$ will be reduced, and vice versa. The value of $\alpha^Q$ reached in the coexisting phase determines the production of isospin-rich meson resonances and subsequent decays in the hadronization process.  Simultaneously,
the value of $\alpha^H$ of the nuclear component  determines the emission of neutron-rich clusters in the collision.
Therefore, the ratio of isospin relevant meson ({e.g., $\pi^- /
\pi^+$) and neutron-rich cluster production have an opposite tendency when the mixed phase forms in experiments. 
In the hadron-PNJL model, the emission of neutron-rich cluster is reduced because of the larger isospin trapping (larger $\alpha^Q$) in the quark
component of the mixed phase. But in the hadron-$\mu$PNJL model with $\beta>0.4$, the isospin asymmetry of $u,\,d$ quark decreases greatly at high temperature. Correspondingly,  the neutron-rich clusters will increase in comparison with the case of hadron-PNJL model. In fact,  in the two-phase model, only the results given by $\beta>0.4$ fulfill the requirement that the phase transition is from nuclear matter to deconfined quark matter as discussed in Sec.~III~B. Therefore, an combination of the isospin relevant observables including both isospin mesons and neutron-rich clusters in HIC experiments can be used to constrain the value of $\beta$.  
 What's more, a detailed analysis of the generated particles in the transport theory for heavy ion collision is deserving for further study. 

From Fig.~\ref{fig:alfa-T} we can also see that  the value of $\alpha^Q$ is $T$ dependent, which means the strength of isospin relevant signatures depends on the beam energy in experiments.  To look for the location of critical end point and the critical behaviors, the second stage  of the beam energy scan (BES II) will be performed on RHIC  soon. Relevant experiments at intermediate densities is also in plan on NICA/FAIR/J-PACK. In particular, on J-PACK experiments will focus on the low-$T$ and high-density region.
Therefore, the beam energy scan will be performed in a wide energy range and  the relevant signals  can be measured in the next generation facilities.  Through mapping the isospin effects of generated hadrons it provides an optional method to explore  the  transition boundaries from nuclear to quark matter.

We also note that the vector interactions between quarks are not included in this study. We have discussed elaborately the role of vector
interactions on the phase transition in our previous research
\cite{Shao12}. The calculation shows that with the inclusion of isoscalar-vector interaction the transition will move towards higher densities (chemical potential). Then the asymmetry parameter $\alpha^Q$ in the mixed phase will be enhanced due to the enlargement of the imbalance of symmetry energy in the two phases. For more details, one can refer to \cite{Shao11-2, Shao2015}.

\section{summary}
We have studied the properties of quark matter in the improved PNJL model with the chemical potential dependent Polyakov loop potential which  reflects to some degree the back-reaction of matter sector to glue sector.  Compared with the original PNJL model, a superiority of the $\mu$PNJL model is that it can effectively describe the confinement-deconfinement transition at low $T$ and high density region.

Furthermore, we constructed the hadron-$\mu$PNJL two-phase model, and use it to explore the equilibrium transition from asymmetric nuclear matter to quark matter. We derived the boundaries of the phase  transition and analysed the isospin-relevant signatures deduced from the two-phase model. Compared with the hadron-PNJL model, the
calculation shows that the transition curves move to lower density (smaller chemical potential)  when the $\mu$-dependent Polyakov loop potential is taken.  Correspondingly, the isospin asymmetry in the quark component decrease at high $T$ for a larger $\beta$, which leads to a reduced  ratio of $\pi^- /
\pi^+$, $K^0/K^+$, but these observables are not sensitive to $\beta$ at low $T$.  In the future, good data on the location of the mixed phase observed in HIC experiments will give valuable information on the $\mu$-dependence of the Polyakov loop potential.
In particular, we suggest to measure the isospin-relevant signatures in the next generation accelerators such as NICA, FAIR and J-PACK, as well as the BES II program on RHIC, which would be helpful to explore the strongly interacting matter. In addition, the research is deserved to be extended to investigate the evolution of protoneutron star, which will be done as a further study.

\begin{acknowledgments}
This work is supported by the National Natural Science Foundation of China under
Grant No. 11305121, the Specialized Research Fund for the Doctoral Program of Higher
Education under Project No. 20130201120046, the Natural Science Basic Research Plan
in Shanxi Province of China (Program No. 2014JQ1012) and the Fundamental Research
Funds for the Central Universities.
\end{acknowledgments}%

%\end{CJK*}
\end{document}